\documentclass[pra,twocolumn,aps]{revtex4}
\usepackage{amssymb}
\usepackage{amsmath}
\usepackage[dvips]{graphicx}

\begin{document}

\title{Conditional control of the quantum states of remote atomic memories\\ for quantum networking}
\author{D. Felinto, C.~W. Chou, J. Laurat, E.~W. Schomburg, H. de
Riedmatten, and H.~J. Kimble}
\affiliation{Norman Bridge Laboratory of Physics 12-33, California Institute of
Technology, Pasadena, California 91125, USA}
\date{\today}
\maketitle

\textbf{Quantum networks hold the promise for revolutionary advances in
information processing with quantum resources distributed over remote
locations via quantum-repeater architectures.
Quantum networks are composed of nodes for storing and processing
quantum states, and of channels for transmitting states between them. The scalability of such networks relies critically on the
ability to perform conditional operations on states stored in
separated quantum memories. Here we
report the first implementation of such conditional control of two atomic
memories, located in distinct apparatuses, which results in a 28-fold
increase of the probability of simultaneously obtaining a pair of
single photons, relative to the case without conditional
control. As a first application, we demonstrate a high degree
of indistinguishability for remotely generated single photons by the
observation of destructive interference of their wavepackets. Our results
demonstrate experimentally a basic principle for enabling scalable quantum
networks, with applications as well to linear optics quantum
computation.}

In recent years, it has been established that quantum memory is an
essential component for the distribution of entanglement over
arbitrarily long distances using quantum repeaters
~\cite{zoller05,briegel98}. In the quantum repeater protocol,
entanglement is distributed by entanglement swapping through a
chain of spatially-separated entangled pairs of particles. Without
memory, all pairs need to be entangled at the same time for the
entanglement distribution to succeed, an event whose probability
decreases exponentially as the length of the chain increases. On
the other hand, if it is possible to store the entanglement in
spatially-separated quantum memories and if one has a trigger that
unambiguously heralds the entanglement once it is achieved
\cite{chou05}, it is possible to build up the chain of entangled
pairs by entangling different parts of the chain at different
times. By conditioning the evolution of the whole system to the
output of its different parts, an exponential enhancement is
attained in the probability of success of the protocol, which
leads, for example, to the possibility of scalable long-distance
quantum communication~\cite{duan01}. Let us note also that Linear
Optical Quantum Computing ~\cite{knill01,franson04,nielsen04} or
quantum state engineering, with schemes working in an iterative
manner~\cite{fiurasek03}, are both damped exponentially by low
success probability due to the present lack of synchronized single
photon sources. Real-time control for synchronization of many
sources is then a promising way to boost the probability of
simultaneous generation of many target quantum states, and thereby
to enable the practical realization of elaborate procedures.

Since photons are the basic carriers of information over long distances, the
necessity of using memory for communication implies the need to control the
exchange of quantum information between matter (stationary qubits) and light
(flying qubits). Great progress has been achieved recently in this direction
for different systems that could work as single nodes of a distributed
quantum network~\cite{cirac97}. For example, generation of photon pairs~\cite%
{kuzmich03,balic05,thompson06}, storage of single photons~\cite%
{Eisaman05,Chaneliere05}, and high-efficiency retrieval of stored single
excitations~\cite{laurat06} were implemented with atomic ensembles.
Entanglement between atoms and spontaneously emitted light was demonstrated
with both single trapped atoms~\cite{blinov04,volz06} and atomic ensembles~%
\cite{Matsukevich05,deRiedmatten06}. Heralded entanglement between two cold
atomic ensembles by means of a single stored excitation was demonstrated
last year~\cite{chou05}, followed more recently by the achievement of
probabilistic (\textit{a posteriori}) entanglement between two excitations
stored in different ensembles~\cite{Matsukevich06}. In the continuous
variables regime, deterministic entanglement has been obtained between two
vapor cells at room temperature~\cite{julsgaard01}.

A key point that has not been experimentally addressed up to now, however,
is the extent to which these different systems and techniques enable
scalable quantum networks. Only recently three groups reported the use of
feedback to enhance the probability of generating a photon using a heralded
single photon source with memory~\cite{deRiedmatten06,Matsukevich06b,chen06}%
. In the present work, we report the first implementation of
real-time conditional control of two distant quantum memories. The
memory nodes consist of ensembles of cold atoms that can each
store a single collective excitation in a probabilistic, but
heralded, way~\cite{duan01}. Since this excitation can be
transferred with high efficiency to a light field in the
single-photon regime~\cite{laurat06}, the system functions as a
heralded single-photon source, ideal for applications to quantum
repeaters. The conditional control allows us to store an
excitation in one ensemble, while waiting for a trigger signal
indicating the presence of an excitation in the other ensemble.
Relative to operation without this conditional control, we attain
a factor of 28 increase in the probability to generate
simultaneously a single photon from each system. Further increase
is currently limited by the finite coherence time of our
memory~\cite{felinto05,deRiedmatten06}.

\begin{figure*}[htpb!]
\hspace{0.6cm} \includegraphics[width=11.5cm,angle=270]{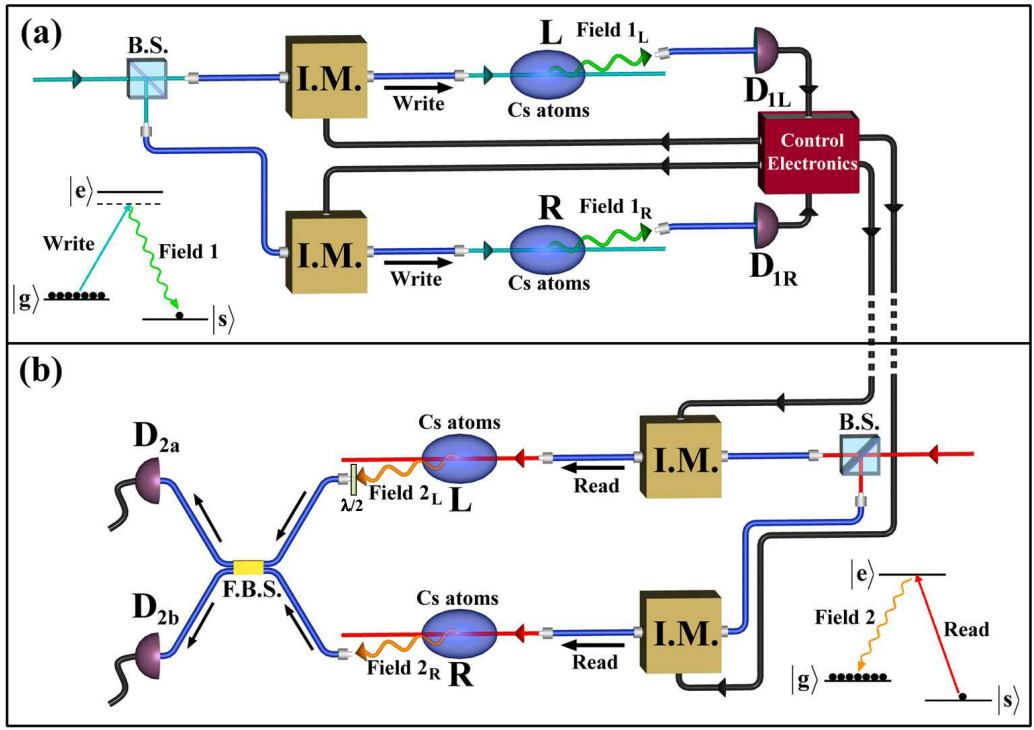} \vspace{%
-0.7cm} \caption{\textbf{Overview of the experiment}. \textbf{a},
Setup for simultaneous storage of single collective atomic
excitations in two ensembles ($L$ and $R$). A write pulse
initially couples the $g\rightarrow e$ transition, resulting with
small probability in the emission of a photon (field 1) on the
$e\rightarrow s$ transition.
Fields $1_L$,$1_R$ are then coupled to polarization-maintaining (PM) fibers and directed to detectors $%
D_{1L}$ and $D_{1R}$, respectively. In a given trial, if neither
of these detectors registers an event, the system is taken back to
its original condition by means of a strong read pulse coupling
the $s \rightarrow e$ transition. On the other hand, if one of the
detectors registers an event, the control electronics uses the
Mach-Zehnder intensity modulators (I.M.) in both write and read
(panel b) pathways to turn off the write-read sequence for the
corresponding ensemble. Later, if the other detector also fires,
the control electronics generates a \textit{ready} signal,
indicating the simultaneous storage of two excitations.
\textbf{b}, Setup for simultaneous generation of two single
photons. The \textit{ready} signal activates the read intensity
modulators, releasing simultaneously the read pulses for both
ensembles. As a result, two photons (in fields $2_L$,$2_R$) are
generated, and then coupled to the two inputs of a PM fiber beam
splitter, whose outputs are directed to detectors
$D_{2a}$,$D_{2b}$. A rotatable half-wave plate ($\lambda/2$)
controls the polarization of field $2_L$.} \label{setup}
\end{figure*}

For applications in quantum repeaters, it is crucial that single photons
emitted by two different ensembles are indistinguishable when combined at a
beam splitter, since this is at the heart of the technique to achieve
measurement-induced entanglement~\cite{duan01,simon03,chou05} and
entanglement swapping~\cite{duan01} between remote atomic systems. As a
first application of our control system, we perform a time-resolved
two-photon interference measurement to quantify the indistinguishability of
the generated single photons, obtained simultaneously from atomic
excitations stored for different amounts of time. We combine the photons at
a beam splitter and observe a nonclassical suppression in the rate of joint
detections at its two output ports~\cite{mandel-wolf95}%
. This suppression is the result of a destructive two-photon interference,
as first demonstrated in a parametric down conversion system by Hong, Ou,
and Mandel~\cite{hong87}. More recently such suppression was observed
for photons emitted successively by a single source~\cite{santori02,legero04}
or simultaneously by different sources, such as spatially separated
down-converters~\cite{deRiedmatten03} and trapped atoms~\cite{beugnon06}.
Our results show a suppression of $(77\pm 6)$\% for the probability of
having two photons leave the beam splitter through different ports, from
which we infer an overlap $\xi \simeq 0.90$ for the wavepackets of the two
single photons.

Our experimental setup is sketched in Fig.~\ref{setup} (see also Methods
section). The two ensembles consist of pencil-shaped clouds of cold cesium atoms located
in two different vacuum chambers, 2.7~m apart. In the beginning of each
trial, all atoms are optically pumped to the hyperfine ground state $%
|g\rangle$ (see Fig.~\ref{setup}a). The atomic level structure for the writing
process consists of the initial ground state $|g\rangle $ ($6S_{1/2},F=4 $
level of atomic cesium), the ground state $|s\rangle $ for storing a
collective spin flip ($6S_{1/2},F=3$), and the excited level $|e\rangle $ ($%
6P_{3/2},F=4$). Then a weak write pulse, lasting 38~ns,
excites the $g \rightarrow e$ transition. With a small probability $%
q_{1}\simeq 0.005\ll 1$, the atomic ensemble spontaneously emits a photon
(field 1) on the $e \rightarrow s$ transition, into the solid angle of our
detection system. For our experimental conditions, the detection of this
first photon heralds the storage of an excitation in a collective, symmetric
mode of the whole ensemble~\cite{duan01,laurat06,duan02}.

This collective excitation can then be retrieved with high probability~\cite%
{duan01,laurat06} by a strong read pulse (38~ns long and resonant to the $%
s\rightarrow e$ transition) counterpropagating with respect to the
write beam; see Fig.~\ref{setup}b. The read pulse results in the
generation of a second photon (field 2) in the direction opposite
to field 1 \cite{balic05}. For both ensembles, a detection in
field 1 occurs with probability $p_{1}=0.12\%$ and
is followed by a detection in field 2 with conditional probability $%
p_{c}\simeq 8.5\%$, which corresponds, after taking the losses in
the field-2 channels into account, to $q_{c}\simeq 34\%$ retrieval
efficiency for the collective mode at the output of the
ensemble~\cite{laurat06}. If no detection in field 1 is registered
in a given trial, the read pulse is fired in order to optically
pump the atoms back to their initial condition. For the chosen
$p_1$, the normalized intensity cross-correlation function between
fields 1 and 2~is measured to be approximately $g_{12}\simeq 23$
for both ensembles \cite{kuzmich03,laurat06}, corresponding to a
field 2 well within the single-photon regime, with a large
suppression of its two-photon component~\cite{laurat06}. For our
system, $g_{12}>2$ is already a strong indication of nonclassical
correlations between fields 1 and 2~\cite{kuzmich03}. The
parameter $w$ quantifies this suppression by examining the
probability of generating two photons in the same pulse,
normalized by an equally bright Poisson-distributed source
\cite{chou04}. Classical fields must satisfy the Cauchy-Schwarz
inequality $w \ge 1$; for independent coherent states, $w=1$,
while for thermal fields, $w=2$~\cite{chou04}. The parametric
dependence of $w$
on $g_{12}$ for our system was investigated in detail in Refs.~%
\onlinecite{laurat06} and~\onlinecite{chou04}, from which we infer
$w\approx 0.17$ for $g_{12}=23$.

Fields $1_{L}$,$1_{R}$ are guided to the respective detectors ($D_{1L}$,$%
D_{1R}$) (Fig.~\ref{setup}a). The field-2 outputs of the two
ensembles, on the other hand, are combined at a fiber beam
splitter, whose outputs are then directed to two detectors
$D_{2a}$ and $D_{2b}$ (Fig.~\ref{setup}b). To exploit the quantum
memory to speed up probabilistic quantum protocols that require
concurrent state preparation in two atomic ensembles, we have
designed and implemented a custom logic circuit that allows
conditional control of the writing, storing, and reading
operations for the atomic excitations. Upon receipt of a field-1
detection signal from one ensemble, the circuit gates off the
write and read pulses for that ensemble, thereby storing a
collective excitation in the atoms
\cite{laurat06,deRiedmatten06,chou04}. The write-read pulse train
in the other ensemble is not affected. The storage stops and the
excitations in the ensembles are read out when either of the
following events occurs: (1) A field-1 photon is detected in the
second system, prompting the circuit to release read pulses into
the first and second ensembles, thereby simultaneously retrieving
the stored excitations from both ensembles; (2) A pre-determined
maximal storage time $\Delta t_{\max }$ set in the circuit is
reached in the first ensemble. Then, the storage ceases, and the
ensembles are optically pumped back to the original condition. The
logic circuit returns to its dormant state, passing all the write
and read pulses to the ensembles, until the next field-1 detection
signal triggers its function.

\vspace{1.5cm}
\begin{figure}[th]
\centerline{\includegraphics[width=9cm,angle=0]{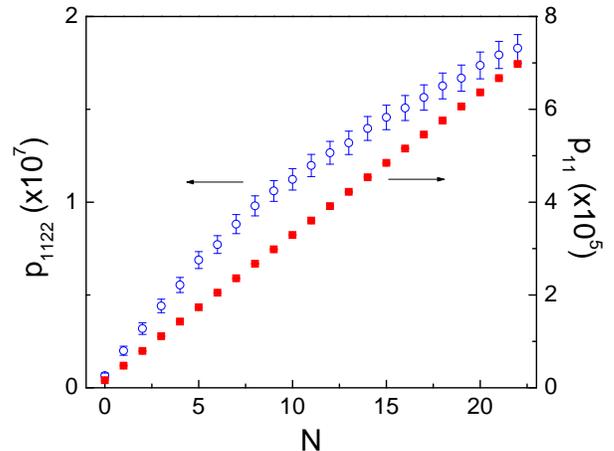}}
\vspace{-2.5cm} \caption{Probabilities $p_{11}$ and $p_{1122}$ of
coincidence detection as functions of the number $N$ of trials
waited between the independent preparations of the two ensembles
$(L,R)$ with $1$ excitation each. Filled squares give the joint
probability $p_{11}$ of simultaneously preparing the two
ensembles. Open circles give the joint probability $p_{1122}$ of
preparing the two ensembles and detecting a pair of photons, one in each output of the beam splitter, in fields $2_{L}$%
,$2_{R}$. Error bars indicate $\pm\sqrt{C}$ photon counting noise,
where $C$ is the number of counts. Total number of counts for the
evaluation of $p_{11}$ was 234244, and 614 for $p_{1122}$. Total
number of trials was $3.36\times 10^{9}$, corresponding to 6 hours
of data taking. The polarizations for fields $2_{L}$,$2_{R}$ were
set to be orthogonal.} \label{p11}
\end{figure}

Our logic circuit is designed specifically to increase $p_{11}$,
the probability of having excitations stored simultaneously in
each ensemble. Given an initial event in one ensemble, the filled
squares in Fig.~\ref{p11} show how $p_{11}$ increases as a
function of the number of trials $N$ of duration $\Delta
t_{trial}=525$~ns that we wait for the other ensemble. The values
in this curve were obtained by performing the experiment using a
maximum number of trials $N_{max}=23$ (corresponding to our
specific $\Delta t_{\max} = 12 \mu$s), recording a file with the
whole history of events, counting the number of events where the
second ensemble was prepared up to $N $ trials after the first
one, and then dividing this number by the total
number of trials $N_{t} $. After $23$ pulses, we observe an increase $%
\mathcal{F} _{11}\simeq 44$ in $p_{11}$, close to the expected value of $%
\mathcal{F} _{11}=(2N_{max}-1)=45$ for the case $p_{1}<<1$ (see
Methods Section). Note that for our experimental conditions, the
conditional logic reduces (with reference to a try-until-success
strategy) the number of trials $N_{t}$ in a negligible way, by up
to $6\%$ (i.e., $2\,p_{1}N_{max}=0.055$). It is also important to
point out that the advantage of such conditional logic will be
(exponentially) more pronounced for larger systems involving more
than two memories.

Our interest, however, is not directly in $p_{11}$, but rather in the
probability $p_{1122}$ of obtaining joint detections for field 2 after the two
systems $(L,R)$ have each stored a single excitation. In Fig.~\ref{p11} $%
p_{1122}$ (open circles) is given as a function of $N$, for the
case where the fields $2_L$,$2_R$ from the two ensembles have
orthogonal polarizations. Evidently, $p_{1122}$ does not increase
by the same factor as does $p_{11}$, but grows instead only up to
$\mathcal{F} _{1122}=28$ [i.e., $28$\ times the value obtained
without the control circuit ($N=0$)]. This behavior is expected
from the fact that the retrieval efficiency for the stored
excitation in the ensemble that is first prepared decays until the
other ensemble is prepared and the read pulses are fired. The
coherence time $\tau
_{c}$ for our system is about 10~$\mu$s~\cite%
{felinto05,deRiedmatten06}, and can also be directly inferred from
the decay with $N$ of the probabilities for one and two
photo-detections in field 2 (see Appendix).

\begin{figure}[tbph]
\centerline{\includegraphics[width=8cm]{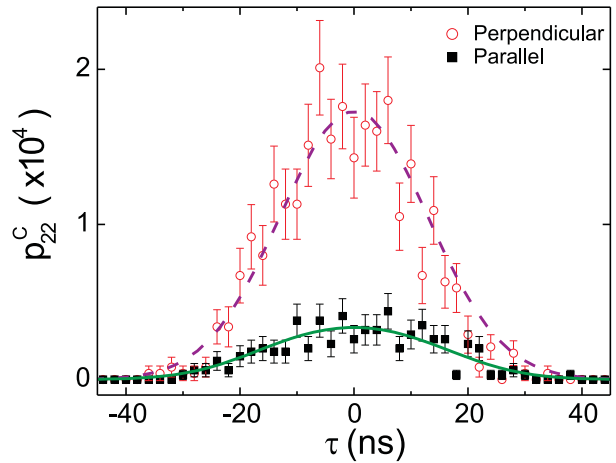}}
\caption{Conditional joint-detection probability
$p_{22}^{c}(\protect\tau)$ of recording events in both $D_{2a}$
and $D_{2b}$, once the two ensembles are ready to fire, as a
function of the time difference $\protect\tau$ between the two
detections. Filled squares (empty circles) provide the results
when the field-2 outputs of the two ensembles are combined with
parallel (orthogonal) polarizations. The dashed line is obtained
from a Gaussian fit
of the orthogonal-polarization data, with half-width at $1/e$ equal to $%
T=18.4\pm 0.2$~ns. The solid line is obtained from the dashed one
by multiplying it by $1-V_{fit} \cos(\Delta \omega \tau)$, with
$V_{fit}=0.80$ and $\Delta \omega/2\pi=4 MHz$. Error bars indicate
$\pm\sqrt{C}$ photon counting noise, where $C$ is the number of
counts.} \label{fitgaussian}
\end{figure}

To characterize the indistinguishability of the photons in fields $2_L$ and $%
2_R$, we perform a measurement of the suppression of
joint-detection events at detectors $D_{2a},D_{2b}$ in the same
trial. In Fig.~\ref{fitgaussian}, we observe that the conditional
joint-detection probability $p_{22}^{\,c}$ decreases for the
situation where fields $2_L$,$2_R$ are combined with the same
polarizations (filled squares), compared to the case of orthogonal
polarizations (open circles). The detection times $t_{d}$ are
obtained from the record of events in our acquisition card, which
has 2~ns resolution, much smaller than the duration of the photon
wavepackets. From this list of detection times, we obtain the time
difference $\tau$ between the two detections. To quantify the
coincidence suppression shown in
Fig.~\ref{fitgaussian}, we consider the visibility $V\equiv(p_{\bot}-p_{\,%
\|})/p_{\bot}$, where $p_{\bot}$ gives $p_{22}^{\,c}$ with orthogonal
polarization and $p_{\,\|}$ gives that with parallel polarization. Note
that $V=1$ for ideal single photons with perfect overlap of their
wavepackets, while $V=0$ for completely distinguishable fields. Obtaining $%
p_{\bot}$ and $p_{\,\|}$ from the integration over $\tau$ of the
respective data points in Fig.~\ref{fitgaussian}, we find
$V=0.77\pm0.06$ (see also Appendix).

For photons orthogonally polarized, we fit our measurements for the
joint-detection probability $p_{22}^{\,c}(\tau )$ in Fig.~\ref{fitgaussian}
to a Gaussian (dashed line),
\begin{equation}
p_{22}^{\,c}(\tau )=p_{0}\,\text{exp}\left( -\frac{\tau ^{2}}{T^{2}}\right)
\text{ .}
\end{equation}%
We find $T=18.4\pm 0.2$~ns, from which the photon duration can be
inferred assuming identical wavepackets. For perfectly
transform-limited photons, the suppressed $p_{22}^{\,c}(\tau )$
with parallel polarizations can be obtained from the one with
perpendicular polarizations by multiplication by a constant
scaling factor $f=1-V_{fit}$. When one introduces frequency jitter
$\Delta \omega$, $p_{22}^{\,c}(\tau )$ can be expressed as
\cite{legero04,beugnon06,legero03}:
\begin{eqnarray}
p^c_{22}(\tau)=\left[p_0\,\text{exp}\left(-\frac{\tau^2}{T^2}\right)\right].\left[1-V_{fit}
\cos(\Delta \omega \tau)\right]
\end{eqnarray}.

The green line gives such a fit with $V_{fit}=0.80\pm 0.02$ and
$\Delta \omega/2\pi=4\pm 4 MHz$. The jitter error bar is obtained
by doubling the $\chi^2$ fitting parameter. This simple fit agrees
well with the data and leads to a time-bandwidth product equal to
$1.2\pm0.2$, providing a good indication that our photons are
close to transform limited. Others measurements are necessary to
investigate this issue further.

The main cause of visibility reduction $V<1$ is that the
two-photon components for the conditional fields $2_{L}$,$2_{R}$
for both ensembles are necessarily nonzero~\cite{laurat06}. From
the inferred value of two-photon suppression $w=0.17$, we estimate
with a simple model that the maximal achievable visibility for
perfect overlap between the two fields is $V_{\max }=0.85$ (see
Appendix). By comparing our measured visibility $V=0.77$ to
$V_{\max }$, we infer that the overlap of the field-2 wavepackets
is $\xi \simeq 0.90$, where $\xi =1$ for perfect mode-matching.
The overlap mismatch can be partly explained by a non-zero
polarization extinction ratio in the polarization-maintaining
fibers (-14 dB). The visibility also decreases due to a small
misalignment introduced by the rotation of the half-wave plate to
switch between the two polarizations. The alignment was checked by
analyzing the events where the two photons are created in
different readout intervals, thus exhibiting no interference (see
Appendix). In this case, the orthogonal-polarization
joint-detection level is 0.08 smaller than for parallel
polarization, which should increase the visibility by 0.02.
Another small decrease in the visibility should come from the
measured imbalance (0.51/0.49) of the fiber beam splitter. Note
that the visibility of the two-photon interference can in
principle be increased by reducing the
intensity of the write pulses (and thereby $q_{1}$ and hence $w$ \cite%
{chou04,laurat06}), but this comes at the expense of a reduced counting rate.

\vspace{2cm}
\begin{figure}[htpb!]
\centerline{\includegraphics[width=9cm,angle=0]{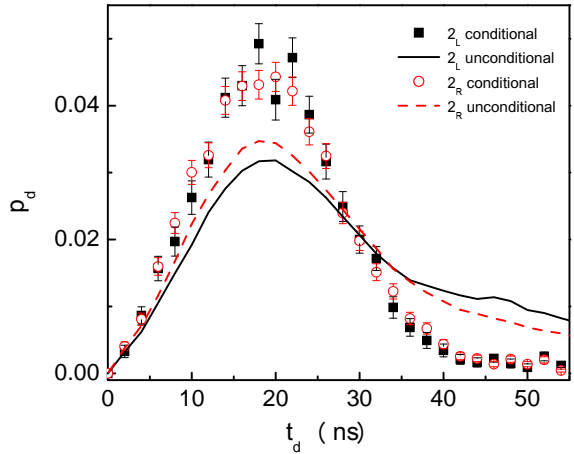}} \vspace{%
-2.5cm} \caption{Probability densities $p_d$ for the wavepackets
of fields $2_L$ and $2_R$ derived from the distribution of
detection events over time $t_d$. Filled squares (empty circles)
give the conditional wavepacket for field $2_L$ ($2_R $), for
detections occurring only in the same trial as a field-1
detection.
The solid (dashed) line gives the unconditional wavepacket for ensemble $L$ (%
$R$). The data-taking duration was 200~s for ensemble $L$, and
300~s for ensemble $R$. All curves are normalized by their areas.
Error bars indicate statistical errors.} \label{vis}
\end{figure}

We also performed independent measurements to assess the temporal overlap of
the two wavepackets. The temporal shape of the conditional wavepackets of
fields $2_{L}$,$2_{R}$ are given, respectively, by the filled squares and
empty circles in Fig.~\ref{vis}. These wavepackets were obtained by blocking
the fields from the other ensemble, disabling the logic circuit, and
considering only the field 2 detections occuring in the same trial as a
field 1 detection. The two wavepackets are quite similar, consistent with
the good suppression we measured, and both have temporal widths around $%
T_{c}\simeq 13$~ns (Gaussian fit), which is also consistent with
the expected width $T/\sqrt{2}=13\pm0.2$ns extracted from
Fig.~\ref{fitgaussian} by considering identical wavepackets. In
Fig.~\ref{vis} we also show the unconditional wavepackets for
fields $2_L$,$2_R$. They are considerably different from the
conditional ones, indicating that part of the light we collect
from the ensembles is not correlated to field 1, but likely arises
from processes not related to the collective atomic
state~\cite{duan02}.

In summary, we have demonstrated a $28$-fold enhancement in the
probability for simultaneous generation of single photons from two
remote atomic ensembles, as a result of the real-time control of
the storage of collective excitations in the ensembles. We have
used this enhancement to demonstrate the indistinguishability of
two photons generated by independent systems, through the
observation of a destructive interference of their wavepackets
that results in the coalescence of the two photons at a beam
splitter. From this measurement, we have also inferred that the
generated single photons are narrowband, and their wavepackets are
close to transform-limited. We emphasize that absent conditional
control of the two remote systems, or without quantum memory, data
of comparable quality to that presented here would have required
continuous acquisition over more than two weeks, which is
prohibitive. The fundamental control for quantum state
manipulation implemented here will be integral to future advances
with networks of quantum memories, including for quantum
repeaters. Our work thereby paves the way to scalable quantum
networks over distances much longer than set by fiber optic
attenuation.

\vspace{0.2cm} \noindent {METHODS}

\noindent \textbf{Experimental details.} Magneto-optical traps are
used to form the clouds of atoms, and are switched off for 6~ms
every 25~ms period. After waiting for the trap magnetic field to
decay~\cite{felinto05}, a train of write and read pulses excite
the sample during the last 2~ms. The write pulse is 10~MHz red
detuned from the $g\rightarrow e$ transition. The
transverse waist of the write beam is 200~$\mu $m, and its peak power $%
P_{write}\approx 2\,\mu $W. We collect the light emitted by the
ensemble in a polarization-maintaining (PM) fiber, whose projected
mode on the ensemble corresponds to a beam with 50~$\mu $m waist
intersecting the write-beam direction at a three-degree
angle~\cite{laurat06}. In the experiment, the read pulse is
delayed from the write pulse by 300 ns, leaving time for the
pulses to be gated off after the heralding signal, which occurs
100 ns after the write pulse due to propagation delays.

\noindent \textbf{Increase in probability.} Assume that $p_{1}$ gives the
probability per trial of storing a collective excitation, and that it is
possible to wait up to $N$ trials before reading out the excitation and
releasing the corresponding single photon. The probability of having two
ensembles storing excitations in the same trial is then
\begin{align}
p_{11}=& \,p_{1}\Big\{p_{1}+2\left[ (1-p_{1})p_{1}+(1-p_{1})^{2}p_{1}\right.
\notag \\[0.0cm]
& \left. \;\;\;\;\;\;+\cdots +(1-p_{1})^{N-1}p_{1}\right] \Big\}  \notag \\%
[0.0cm]
\approx & \,(2N-1)p_{1}^{2}\;,\;\mbox{when}\;p_{1}<<1\,.  \notag
\end{align}
The factor of two in the above expression accounts for the two possible
orders in which the ensembles can be prepared.

\textbf{Acknowledgement -} This research is supported by the
Disruptive Technologies Office (DTO) and by the National Science
Foundation. J.L. aknowledges financial support from the European
Union (Marie Curie fellowship). D.F. acknowledges financial
support by CNPq (Brazilian agency).

\appendix
\section{Decoherence}

\label{Decoherence}

\begin{figure}[tbh]
\centerline{\includegraphics[width=9.0cm,angle=0]{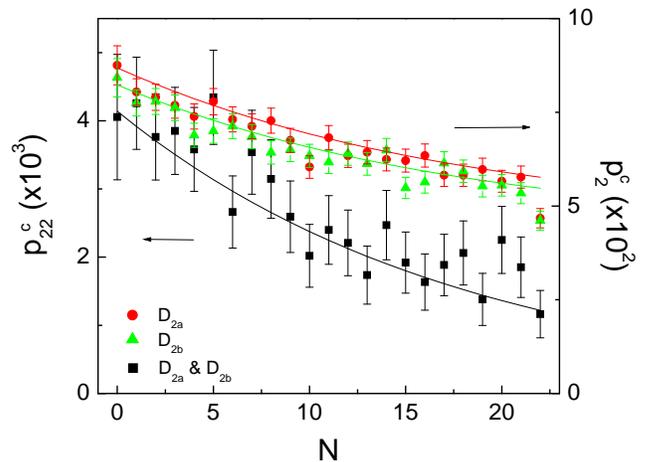}} \vspace{-2.4cm%
} \caption{Conditional probabilities $p_{2}^{c}$ and $p_{22}^{c}$
of measuring one (red circles and green triangles) and two (black
squares) photons in field 2, respectively, once the two ensembles
are ready to fire. The red and black curves are fits using
Eqs.~\eqref{p22c} and~\eqref{p2c}, as discussed in the text. The
green line is $0.95\times $ the red line, as the field 2 level measured by $%
D_{2b}$ is always 5\% lower than for $D_{2a}$, indicating a
possible difference of detection efficiency of this magnitude.}
\label{pc1122}
\end{figure}

Figure~\ref{pc1122} shows the variation of the conditional
probabilities of detecting one ($p_{2}^{\,c}$) and two
($p_{22}^{\,c}$) photons in fields $2_L$ and $2_R$, once the two
ensembles are ready to fire, as functions of the number $N$ of
trials that occur between the two field-1 detections. This figure
was obtained from the same raw data as Fig.~\ref{p11}.
Fields $2_L$ and $2_R$ have then orthogonal polarizations, and are
combined at a beam splitter as shown in Fig.~\ref{setup}b. In order to obtain the quantities in
Fig.~\ref{pc1122}, we divided the number of coincidences in field
2 that followed two temporally separated detections in field 1 by
the number of times the two ensembles were prepared with that
specific time separation. The solid lines are fittings considering
an exponential decay of the conditional probability $p_{c}$ for
the second photon from
either of the two ensembles, once a detection has occurred in field 1. We assumed the same $%
p_{c}$ and decay time for the two systems. Note that the two
systems were actually set up to have similar $p_{c}$ and similar
Raman
linewidths for transitions between the hyperfine ground states~\cite%
{felinto05}, which should correspond to the system coherence time.
The expressions used to fit were then

\begin{equation}
p_{22}^{\,c}(N)=\frac{p_{c}^{2}}{2}e^{-N/Nc}\;,  \label{p22c}
\end{equation}%
\begin{equation}
p_{2}^{\,c}(N)=\frac{p_{c}+p_{c}e^{-N/Nc}}{2}-p_{22}^{\,c}(N)\;.
\label{p2c}
\end{equation}

\noindent We assume above that $p_{22}^{\,c}$ always involves one
photon coming from an excitation stored during $N$ trials. In this
way, we are neglecting the two-photon component of field 2, as
well as diverse sources of background. For $p_{2}^{\,c}$, the
first two terms take into account that the conditioned detection
event can be originated from either an ensemble that has just been
excited, or a stored (for $N$ trials) excitation. The third term
subtracts the probability of having a joint detection on field 2
($p_{2}^{\,c}$ gives the probability of detecting an event on one
detector and zero on the other). From the fitting, we then obtain
$p_{c}=0.091$ and $N_{c}=18$, corresponding to a coherence time
$\tau _{c}=N_{c}\times 0.525\mu \mathrm{s}=9.5 \,\mu \mathrm{s}$.
Keeping in mind the simplicity of the above expressions, which do
not take into account any background in field 2 or its two-photon
component, the inferred conditional probability $p_{c}$ is then
consistent with the independently measured values of about 0.085
for each ensemble.

From the above discussion, it is then straightforward to obtain an
expression taking into account decoherence for the measured
$p_{1122}$, presented in Fig.~\ref{p11}. Note first
that, in the ideal case of very long coherence time ($N_c
\rightarrow \infty$), $p_{1122}$ can be obtained from the $p_{11}$
expression presented in the Methods section by
multipling it by the conditional probability $p_{22}^{\,c} (0)$ of
obtaining a pair of photons with effectively zero delay ($N/N_c
\rightarrow 0$) between them:
\begin{equation}
p_{1122}^{ideal} \approx \frac{(2N-1)
p_1^2p_c^2}{2}\;,\;\mbox{when}\;p_{1}<<1\,. \label{p1122ideal}
\end{equation}

\vspace{1.6cm}

\begin{figure}[htpb!]
\centerline{\includegraphics[width=9.0cm,angle=0]{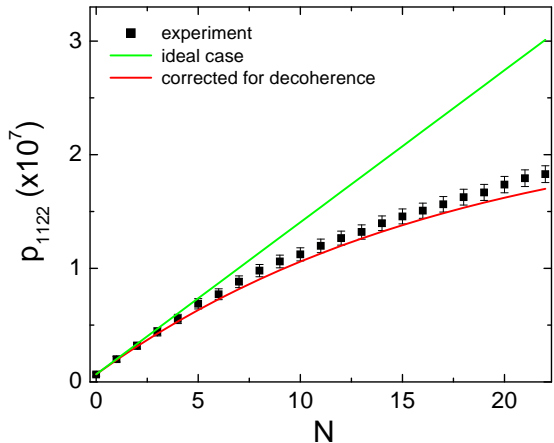}} \vspace{-2.4cm%
} \caption{Probability $p_{1122}$ of coincidence detection as
functions of the number $N$ of trials waited between the
independent preparations of the two ensembles $(L,R)$ with $1$
excitation each. Filled squares give the measured joint
probability $p_{1122}$ of
preparing the two ensembles and detecting a pair of photons, one in each output of the beam splitter, in fields $2_{L}$%
,$2_{R}$. Error bars indicate statistical errors. The
polarizations for fields $2_{L}$,$2_{R}$ were set to be
orthogonal. These experimental results were also presented in
Fig.~\ref{p11}. The green curve gives the
theoretically expected $p_{1122}^{ideal}$ for the ideal case of
very long coherence time, as given by Eq.~\ref{p1122ideal}. The
red curve gives the theoretical $p_{1122}$ for the case of a
finite coherence time, Eq.~\ref{p1122}.} \label{p1122theory}
\end{figure}

\noindent Considering the measured value of $p_1^2$ and the value of $p_c$
obtained from the fits of Fig.~\ref{pc1122}, we then obtain the
green line plotted in Fig.~\ref{p1122theory}. In order to
introduce decoherence in this analysis, each term of the $p_{11}$
expression deduced in the Methods section should be multiplied by
the proper $p_{22}^{\,c} (N)$ as defined in Eq.~\ref{p22c}:
\begin{widetext}
\begin{equation}
p_{1122}(N) = \,p_{1}\Big\{p_{1}p_{22}^{\,c}(0)+2\left[
(1-p_{1})p_{1}p_{22}^{\,c}(1)+(1-p_{1})^{2}p_{1}p_{22}^{\,c}(2)+\cdots
+(1-p_{1})^{N-1}p_{1}p_{22}^{\,c}(N-1)\right] \Big\} \,.
\label{p1122}
\end{equation}
\end{widetext} A plot of this expression, considering the $p_c$ and
$N_c$ obtained from the fits in Fig.~\ref{pc1122}, is shown as the
red curve in Fig.~\ref{p1122theory}. The quite reasonable
agreement with the experimental data (filled squares) indicates
then that the experimentally observed increase in $p_{1122}$ can
be understood by the increase in $p_{11}$ provided by the circuit
combined with the decoherence of the stored collective excitation.

Note finally that $p_{1122}$ times the number of trials per second
gives the rate of conditional joint detections in fields
$2_L,2_R$. In this way, from Fig.~\ref{p1122theory} we can see
that a larger coherence time could still enhance this detection
rate by up to a factor of 1.6 for $N=23$ (enhance the factor
$\mathcal{F} _{1122}$ from 28 to 45). An increase on the
conditional probability $p_c$ would also greatly improve this
rate, since it scales with $p_c^2$. As discussed in the text, we
infer that the probability $q_c$ of extracting the photon from the
ensemble is about $34\%$ for our experimental conditions. Thus,
considering the same amount of losses on the field 2 pathways and
the same detection efficiencies, we infer that ideally, if one
achieved $q_c=1$, $p_{c}$ can still be increased by up to a factor
of 3, which would increase the joint detections rate by 9. This
indicates a good prospect for further optimizations of our system
in the future. \vspace{1.6cm}

\section{Visibility as a function of $w$}

In the case of two indistinguishable single-photon wavepackets
combined at a 50/50 beam splitter (BS), no coincidences should be
observed at the two ouputs of the BS~\cite{mandel-wolf95}.
However, if the combined fields are not perfectly overlapping
single photons, coincident detections at the output of the BS can
occur due to the two-photon component in each input port. In this
section, we evaluate the loss of visibility due to this effect with a simple model.

Suppose that the two ensembles are prepared each with stored
excitation, as heralded by a detection in field 1 for both
ensembles. Let's denote $P_{n}$ the probability of finding $n$
photons in field 2, and assume, for simplicity, the various
$P_{n}$ are the same for both ensembles. In each field (before the
BS), the two-photon suppression is characterized by the parameter
$w$ \cite{chou04}:
\begin{equation}
w=\frac{2P_{2}}{P_{1}^{2}}\,,
\end{equation}%
so that the two-photon component can be written as:
\begin{equation}
P_{2}=\frac{wP_{1}^{2}}{2}\,.
\end{equation}%
Let us now combine the two fields at the BS. The probability to
have one photon at each output of the BS, when the two wavepackets
do not overlap (e.g., if they have orthogonal polarizations) is
given by:

\begin{equation}
p_{\bot }=\frac{P_{2}}{2}+\frac{P_{2}}{2}+\frac{P_{1}^{2}}{2}=\frac{%
wP_{1}^{2}}{4}+\frac{wP_{1}^{2}}{4}+\frac{P_{1}^{2}}{2}\text{ ,}
\label{port}
\end{equation}

 where the two first terms corresponds to the terms with two
photons in one input mode of the BS, and the third term to the
case with one photon in each input mode. The factor 1/2
corresponds to the 50\% chance that the photons split at the BS.
In this simplified calculation, we neglect the case where we have
two photons in one input and one in the other one, whose
probability is on the order of $P_{1}^{3}$.

If the two fields overlap perfectly at the BS (parallel
polarizations), the term with one photon in each input does not
lead to coincidences, and the probability to have one photon in
each output is then:
\begin{equation}
p_{\Vert }=\frac{wP_{1}^{2}}{4}+\frac{wP_{1}^{2}}{4}\,.
\label{ppar}
\end{equation}%
Taking Eqs.~\eqref{port} and~\eqref{ppar} into account, we find
that the
visibility can be written as:%
\begin{equation}
V=\frac{p_{\bot }-p_{\Vert }}{p_{\bot }}=\frac{1}{1+w}\,.
\end{equation}
In our case, we have $g_{12}\approx 23$ for the two ensembles,
from which we estimate $w\approx 0.17$~\cite{laurat06}. This leads
to a maximal visibility of $V_{max}=0.85$ for a perfect overlap
$\xi =1.0$ between the fields. From our measured visibility of
$0.77$, we then estimate an overlap $\xi =0.90$.

\section{Joint-detection levels for events in different trials}

\begin{figure}[th]
\centerline{\includegraphics[width=8.5cm,angle=0]{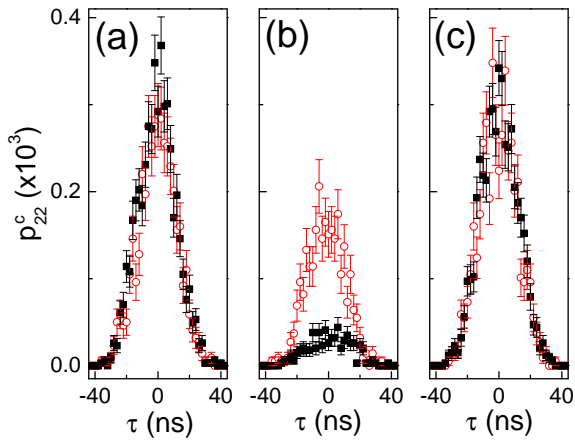}}
\vspace{-2.0cm} 
\caption{Conditional joint-detection probability
and $D_{2b}$, once a ready signal is generated as a result of the
two ensembles being ready to fire, as a function of the time
difference $\protect\tau $\ between the two detections. (b) The
two detections occur within the same trial. (a) Detector $D_{2a}$
fires first and $D_{2b}$ fires after the next ready signal. (c)
Same as (a), but with the detector order inverted. The red circles
(black squares) provide the results for field 2 from the two
ensembles having orthogonal (parallel) polarizations.} \label{dip}
\end{figure}

In Fig.~\ref{dip}b, we show how the conditional probability of
detecting two photons, when the $(L,R)$ systems are ready,
decreases as a function of the
delay between the two detections for the situation where the fields $2_L$,$%
2_R$ are combined with the same (red) or orthogonal (black)
polarizations. It corresponds then to Fig.~\ref{fitgaussian}.
The time $t_{d}$ of the detections is obtained from the recording
of events in our acquisition card, and it refers to a fixed
reference that marks the beginning of the 525~ns repetition
periods. From this list of detection times, we obtain the relative
delay $\tau $ when the two detections occur in the same trial.

\vspace{2.0cm}

We also show in Fig.~\ref{dip}, the cases where the two detections
in $D_{2a}$ and $D_{2b}$ occur in different trials, with the event
in one detector occuring when the two ensembles are ready and the
event in
the other detector occuring the next time the ensembles are ready. Figures~%
\ref{dip}a and~\ref{dip}c give the cases in which one detector or
the other registers an event first. The fact that the signal level
is similar in both cases, with different polarizations, indicates
that there is no large misalignment when the half-wave plate is
turned to switch between the two polarization configurations. Even
though, if we integrate the curves in Figs.~\ref{dip}a
and~\ref{dip}c, the value obtained for orthogonal polarization is
about 0.08 lower than the one obtained for parallel polarization.
We confirmed this value by calculating it also from other
different-trials peaks (for detection events separated by up to 5
ready signals). The curves with different polarizations were
taking at alternate cycles of half-hour data taking, just turning
a single half-wave plate between them. In this way, we believe the
decrease for orthogonal polarizations comes just from a small
misalignment in the fiber input introduced by this operation.

The level of the peak in (b) obtained with orthogonal
polarizations should be half that observed in (a),(c) for the case
where pure single photons arrive at the beam splitter. The
experimental observed ratio $r$ is found to be $r= 0.60\pm 0.05$.
This ratio can be explained by the two-photon component of our
generated state. As previously done in section II, let's denote
$P_1$ and $P_2$ the probabilities of finding respectively one or
two photons in each field 2. For the sake of simplicity, these
probabilities are taken equal for the two ensembles. Including
two-photon events, the probability for coincidence in the two
detectors is given for the center peak (Fig.~2b) by:
\begin{equation}
\frac{1}{2}P_1^2+P_2
\end{equation}
The first term takes into account the two cases where the single
photons are both reflected or transmitted at the beam splitter.
The second one corresponds to the case where two photons arriving
at one input of the beam splitter are split into the two arms.
Higher-order cases, which involve for instance two photons in each
input, or two photon in one input and one in the other, are
neglected.

 \vspace{1.3cm}
\begin{figure}[th]
\centerline{\includegraphics[width=9cm,angle=0]{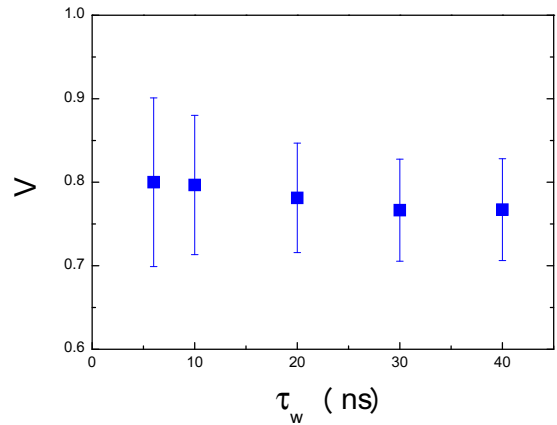}}
\vspace{-3cm} \caption{Visibility $V$ of the two-photon
coincidence suppression as a function of the integration window
for the time difference between the two
field-2 detections. The integration is made from $-\protect\tau _{w}$ to $%
\protect\tau _{w}$.} \label{vis2}
\end{figure}

In a similar way, the probability for the others peaks, where
detections occur in different trials, can be written as
\begin{equation}
\left(P_1+2P_2\right)^2
\end{equation}
This expression corresponds basically to the mean photon number
going to one detector times the mean photon number going to the
other one.

With $P_2=w P_1^2/2$, and by neglecting higher-order terms, the
ratio becomes:
\begin{equation}
r=\frac{1}{2}\frac{1+w}{1+2 w P_1}
\end{equation}
With $P_1=0.085$ and $w=0.17$, this expression gives then an
expected value $r=57\%$, which is consistent with the observed
one.

\section{Visibility as a function of integration windows}
In Fig.~\ref{vis2} we show the results of the measurement of $V$
for different integration windows around $\tau =0$. We see that
for an
integration window around the center, from $\tau=-6$~ns to $\tau=6$%
~ns, the visibility is $80\pm 10\%$, while the integration using
the whole window gives $V=77\pm 6\%$, indicating that the
suppression is roughly uniform for all $\tau$, which is also
consistent with having close to transform-limited wavepackets for
both fields $2_L$,$2_R$.

\section{Time windows}

The electronic time windows used for field 1 and field 2
detections were 80~ns and 90~ns long, respectively, positioned
around the center of the respective wavepackets. In order to
analyze the conditional field-2 wavepackets overlap in Fig.~\ref{fitgaussian}, and also in the above Figs.~\ref{dip}
and~\ref{vis}, we introduced in the analysis an additional time
window, only 44~ns long around the conditional field 2. As can be
seen in Fig.~\ref{vis}, this corresponds to consider
the whole conditional field-2 wavepackets.

\end{document}